\documentclass{article}
\usepackage[utf8]{inputenc}
\usepackage{authblk}
\usepackage{soul}

\usepackage{color}
\usepackage[labelfont=bf]{caption}

\title{Interplay between competitive and cooperative interactions in a three-player pathogen system}

\author[1]{Francesco Pinotti}
\author[2,3,4]{Fakhteh Ghanbarnejad}
\author[2,5]{Philipp H\"{o}vel}

\author[1,*]{Chiara Poletto}

\affil[1]{INSERM, Sorbonne Université, Institut Pierre Louis d’Épidémiologie et de Santé Publique, IPLESP, F75012, Paris, France}
\affil[2]{Institut f\"ur Theoretische Physik, Technische Universit\"at Berlin, Hardenbergstra{\ss}e 36, 10623 Berlin, Germany}
\affil[3]{The Abdus Salam International Centre for Theoretical Physics (ICTP), Trieste, Italy}
\affil[4]{Physics Department, Sharif University of Technology, P.O. Box 11165-9161, Tehran, Iran}
\affil[5]{School of Mathematical Sciences, University College Cork, Western Road, Cork T12 XF62, Ireland}
\affil[*]{chiara.poletto@inserm.fr}
\date{}

\usepackage{amsmath}
\usepackage{graphicx}
\usepackage{color}
\usepackage{pdfpages}

\begin{document}

\maketitle

\begin{abstract}
In ecological systems heterogeneous interactions between pathogens take place simultaneously. This occurs, for instance, when two pathogens cooperate, while at the same time multiple strains of these pathogens co-circulate and compete. Notable examples include the cooperation of HIV with antibiotic-resistant and susceptible strains of tuberculosis, or some respiratory infections with \textit{Streptococcus pneumoniae} strains. Models focusing on competition or cooperation separately fail to describe how these concurrent interactions shape the epidemiology of such diseases. We studied this problem considering two cooperating pathogens, where one pathogen is further structured in two strains. The spreading follows a susceptible-infected-susceptible process and the strains differ in transmissibility and extent of cooperation with the other pathogen. We combined a mean-field stability analysis with stochastic simulations on networks considering both well-mixed and structured populations. We observed the emergence of a complex phase diagram, where the conditions for the less transmissible, but more cooperative strain to dominate are non-trivial, e.g. non-monotonic boundaries and bistability. Coupled with community structure, the presence of the cooperative pathogen enables the co-existence between strains by breaking the spatial symmetry and dynamically creating different ecological niches. These results shed light on ecological mechanisms that may impact the epidemiology of diseases of public health concern.
\end{abstract}

\section{Introduction}

Pathogens do not spread independently. Instead, they are embedded in a larger ecosystem that is characterised by a complex web of interactions among constituent elements. 
Among ecological forces shaping such ecosystems, pathogen-pathogen interactions have drawn increasing attention during recent years 
due to their population-level impact and public health consequences. Recent advances in serological tests and genotyping techniques have improved our reconstruction of pathogen populations where multiple strains co-circulate, often competing due to cross-protection or mutual exclusion. Examples include tuberculosis~\cite{tornheim_tuberculosis_2017,cohen_modeling_2008}, \textit{Plasmodium falciparum}~\cite{chen_competition_2014}, \textit{Streptococcus pneumoniae}~\cite{cobey_niche_2012,opatowski_assessing_2013} and \textit{Staphylococcus aureus}~\cite{obadia_detailed_2015,donker_population_2017}. Polymorphic strains can also interact in more complex ways, with both competition and cooperation acting simultaneously, as observed in co-circulating Dengue serotypes~\cite{reich_interactions_2013}. While interfering with each other, strains also interact with other pathogens co-circulating in the same population. Tuberculosis~\cite{tornheim_tuberculosis_2017}, HPV~\cite{looker_evidence_2018} and \textit{P. falciparum}~\cite{cohen_increased_2005},  
for example, appear to be facilitated by HIV, whereas \textit{S. pneumoniae} benefits from some bacterial infections, e.g. \textit{Moraxella catarrhalis}, and is negatively associated to others such as \textit{S. aureus}~\cite{bosch_viral_2013,dunne_nasopharyngeal_2013}. Competition, cooperation and their co-occurrence may fundamentally alter pathogen persistence and diversity, thus calling for a deep understanding of these forces and their quantitative effects on spreading processes.

Mathematical models represent a powerful tool to assess the validity and impact of mechanistic hypotheses about interactions between pathogens or pathogenic strains~\cite{matt_j._keeling_modeling_2008,shrestha_statistical_2011}. 
The literature on competitive interactions is centered on pathogen dominance and coexistence. 
Several factors were found to affect the ecological outcome of the competition, including co-infection mechanisms~\cite{gjini_slow-fast_2017,sofonea_within-host_2015,alizon_co-infection_2013,alizon_multiple_2013}, host age structure~\cite{martcheva_subthreshold_2007,cobey_host_2017}, contact network~\cite{eames_coexistence_2006, miller_co-circulation_2013,darabi_sahneh_competitive_2014,sanz_dynamics_2014,li_coexistence_2003,chai_competitive_2011,karrer_competing_2011,masuda_multi-state_2006,leventhal_evolution_2015,buckee_effects_2004,pinotti_host_2019,soriano-panos_markovian_2019} and spatial organisation~\cite{poletto_host_2013,poletto_characterising_2015,allen_sis_2004,buckee_host_2007, wu_coexistence_2016,lion_spatial_2016}. At the same time, models investigating cooperative interactions have driven many research efforts
during recent years~\cite{cai_avalanche_2015, hebert-dufresne_complex_2015, lee_universal_2017, cui_mutually_2017, chen_fundamental_2017, rodriguez_diversity_2018,rodriguez_risk_2017}. Cooperation has been found to trigger abrupt transitions between disease extinction and large scale outbreaks along with hysteresis phenomena where the eradication threshold is lower than the epidemic threshold~\cite{cai_avalanche_2015, chen_fundamental_2017,hebert-dufresne_complex_2015}. These findings were related to the high burden of synergistic infections, e.g. the HIV and tuberculosis co-circulation in many parts of the world. 
Despite considerable mathematical and computationally-heavy research on interacting pathogens, competition and cooperation have been studied mostly separately. 
Nevertheless, current understandings about these mechanisms taken in isolation may fail to describe the dynamics arising from their joint interplay, 
where heterogeneous interactions may shape the phase diagram of co-existence/dominance outcome, along with the epidemic prevalence.

Here we studied the simplest possible epidemic situation where these heterogeneous effects are at play. We introduced a three-player model where two pathogens cooperate, and one of the two is structured in two mutually exclusive strains. This mimics a common situation, where e.g. resistant and susceptible strains of \textit{S. pneumoniae} cooperate with other respiratory infections~\cite{bosch_viral_2013}, and allows us to address two important ecological questions: 
\begin{itemize}
  \item How does the interplay between two distinct epidemiological traits, i.e. the transmissibility and the ability to exploit the synergistic pathogen, affect the spreading dynamics? 
  \item How does the presence of a synergistic infection alter the co-existence between competing strains? 
\end{itemize}
We addressed these questions by providing a characterisation of the phase space of dynamical regimes. We tested different modelling frameworks (continuous and deterministic vs. discrete and stochastic), and compared two assumptions regarding population mixing, i.e. homogeneous vs. community structure. 

The paper is structured as follows: Section~\ref{sec:model} introduces the main aspects of the three-player model. We provide the results of the deterministic dynamical equations in Section~\ref{sec:results_mixed}, where we present the stability analysis, together with the numerical integration of the equations, to characterise the phase space of the dynamics. The structuring of the population in two communities is analysed in Section~\ref{sec:continuous-communities}. In Section~\ref{sec:results_networks} we describe the results obtained within a network framework comparing stochastic simulations in an Erd\H{o}s-Rényi and a random modular network. We discuss the implications of our results in Section~\ref{sec:discussion}.

\section{The model}
\label{sec:model}
\begin{figure}[ht!]
  \centering
  \includegraphics[width=\textwidth, keepaspectratio]{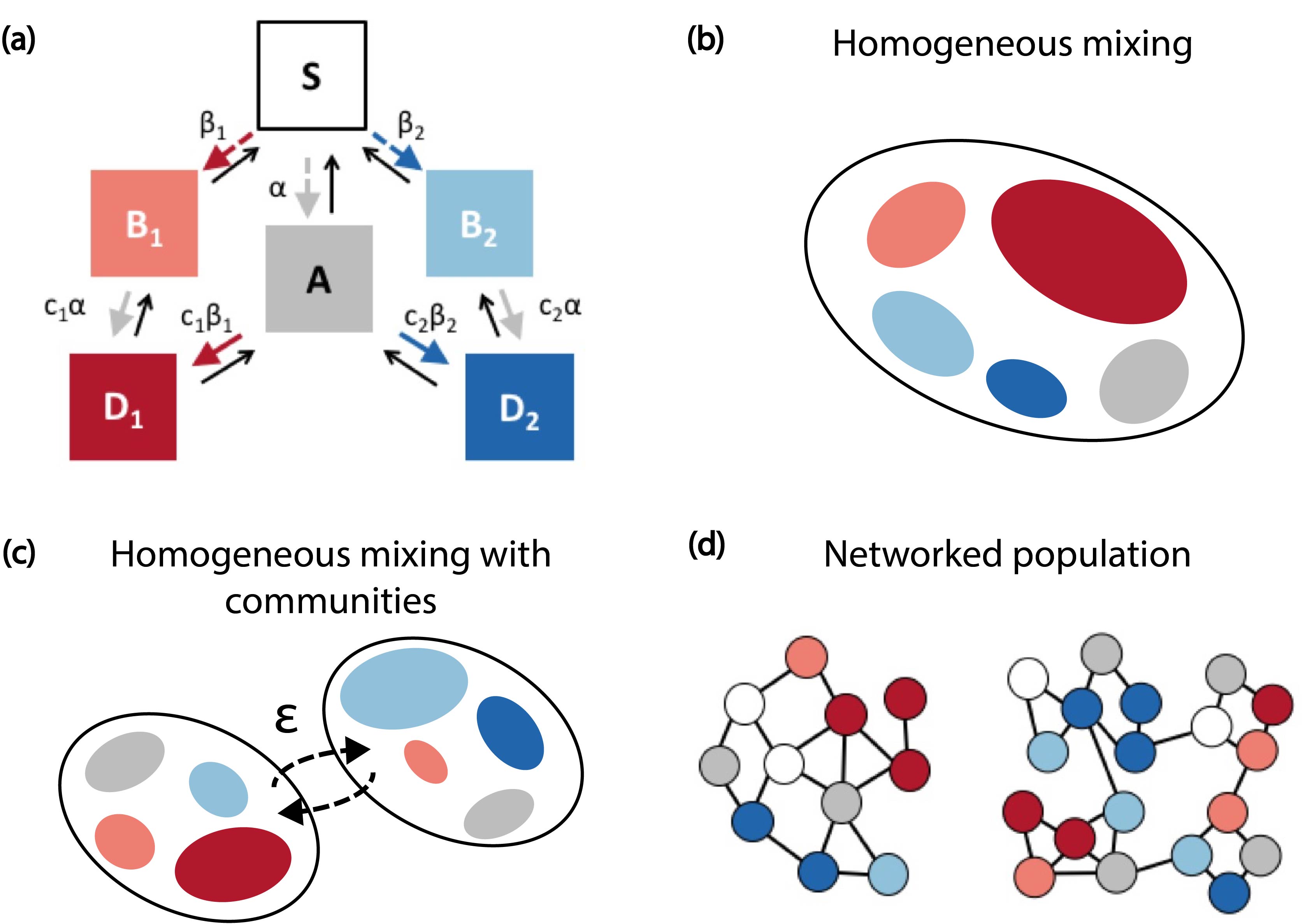}
  \caption{\textbf{Scheme of the model}. (a) Compartmental model. Coloured arrows represent transitions occurring due to infection transmission. Dashed arrows refer to primary infections, while solid arrows refer to secondary ones; transmission parameters are also reported close to each arrow. Black arrows represent recovery transitions. (b-d) Schematic representation of the modelling frameworks and population structures considered. (b) A homogeneously-mixed population (Section~\ref{sec:results_mixed}). (c) Two homogeneous populations with across-group mixing ruled by the parameter $\epsilon$ (Section~\ref{sec:continuous-communities}); in (b),(c), colours indicate the infectious density for each compartment. (d) Erd\H{o}s-Rényi and random modular networks (Section~\ref{sec:results_networks}). Colours indicate the nodes' status.}
  \label{fig:Fig1}
\end{figure}

A scheme of the model is depicted in figure~\ref{fig:Fig1}a. We considered the case in which two pathogens, $A$ and $B$, follow susceptible-infected-susceptible dynamics, and we made the simplification that they both have the same recovery rate $\mu$. $A$ and $B$ cooperate in a symmetric way through increased susceptibility, i.e. a primary infection by one of the two increases the susceptibility to a secondary infection by the other pathogen. We assumed that the cooperative interaction does not affect infectivity, thus doubly infected individuals, i.e. infected with both $A$ and $B$, transmit both diseases at their respective infection rates. $B$ is structured in two strains, $B_1$ and $B_2$, that compete through mutual exclusion (co-infection with $B_1$ and $B_2$ is impossible) and differ in epidemiological traits. Specifically, we denoted the infection rates for pathogens $A$ and $B_i$ with $\alpha$ and $\beta_i$ ($i=1,2$), respectively. We introduced the parameters $c_i>1$ to represent the increased susceptibility after a primary infection. In summary, individuals can be in either one of 6 states: susceptible ($S$), singly infected ($A$, $B_i$) and doubly infected with both $A$ and $B_i$. The latter status is denoted by $D_i$. 

To simplify the analytical expressions we rescaled time by the average infectious period $\mu^{-1}$, which leads to non-dimensional equations. The basic reproductive ratios of the each player, $R^{(i)}_0= \beta_i/\mu$ and $R^{(A)}_0= \alpha/\mu$, become then equal to the transmission rates $\beta_i$ and $\alpha$, respectively. This implies that the threshold condition $\beta_i, \alpha>1$ has to be satisfied in order for the respective player to be able to individually reach an endemic state.
Assuming a homogeneously mixed population, the mean-field equations describing the spreading dynamics are:
\begin{align}
\begin{cases}
\dot{S} & = A + B_1 + B_2 -\alpha S\,X_A -\beta_1S\,X_1 - \beta_2S\,X_2 \\
\dot{B}_1 & = D_1-B_1-c_1\alpha B_1\,X_A + \beta_1 S\,X_1 \\
\dot{B}_2 & = D_2-B_2-c_2\alpha B_2\,X_A + \beta_2 S\,X_2 \\
\dot{A} & = D_1 + D_2 - A + \alpha S
\,X_A - c_1\beta_1 A\,X_1 - c_2\beta_2 A\,X_2 \\
\dot{D}_1 & = -2D_1 + c_1\alpha B_1\,X_A + c_1\beta_1 A\,X_1 \\
\dot{D}_2 & = -2 D_2 + c_2 \alpha B_2\,X_A + c_2\beta_2 A\,X_2 ,
\label{Eq:system}
\end{cases}
\end{align}
\noindent where the dot indicates a differentiation with respect to time rescaled by $\mu^{-1}$, and quantities $S$, $A$, $B_i$ and $D_i$ represent occupation numbers of the compartments divided by the population. 
The variables $X_A$, $X_i,\: i=1,2$, indicate the total fractions of individuals carrying $A$ and $B_i$, respectively, among the singly and doubly infected individuals. They satisfy the equations:
\begin{subequations}
\begin{align}
        \dot{X}_i & = X_i \beta_i(S + c_i\,A) - X_i ,  \label{Eq:Xi}\\
        \dot{X}_A & = X_A \alpha(S + c_1 B_1 + c_2 B_2) - X_A. \label{Eq:XA}
\end{align}
\end{subequations}

Without loss of generality, we considered the case in which the strain $B_2$ is more transmissible than $B_1$, i.e. $\delta_{\beta} = \beta_2 - \beta_1 > 0$. Furthermore, we focused on the more interesting case of trade-off between transmissibility and cooperation to limit the parameter exploration: The less transmissible strain, $B_1$, is more cooperative, $\delta_{c} = c_1 - c_2 > 0$. If $B_2$ is more cooperative, we expect it to win the competition. To summarize, our main assumptions are:
\begin{itemize}
    \item $\delta_{\beta} = \beta_2 - \beta_1 > 0$,
    \item $\delta_{c} = c_1 - c_2 > 0$,
    \item $c_i > 1 \: \: \: i=1,2$.
\end{itemize}

In the Results section we will first describe the dynamics arising from the deterministic equations~\eqref{Eq:system}. We will then consider the case in which the whole population is structured in two groups (see figure~\ref{fig:Fig1}c). Finally, we will apply the proposed model to contact networks, where nodes represent individuals and transmission occurs through links, and consider transmission and recovery as stochastic processes. Two types of networks will be tested: Erd\H{o}s-Rényi and random modular networks (see figure~\ref{fig:Fig1}d).

\section{Results} 
\label{sec:results}
\subsection{Continuous well-mixed system}
\label{sec:results_mixed}
We carried out a stability analysis to classify the outcome of the interaction as a function of the difference in strain epidemiological traits, $\delta_{c}$ and $\delta_{\beta}$. Specifically we computed explicit analytical expressions for states' feasibility and stability conditions in several cases. Furthermore, we performed extensive numerical simulations in cases where closed expressions were difficult to obtain. We present the overall behaviour and the main analytical results in this section and we refer to the Supplementary Material for the detailed calculations. In the following, we will use square brackets to indicate final state configurations in terms of persisting strains, thus $[A\&B_1]$ indicates, for instance, the equilibrium configuration where both $A$ and $B_1$ persist, while $B_2$ becomes extinct.

\begin{figure}[ht!]
\centering
\includegraphics[width=\textwidth,keepaspectratio]{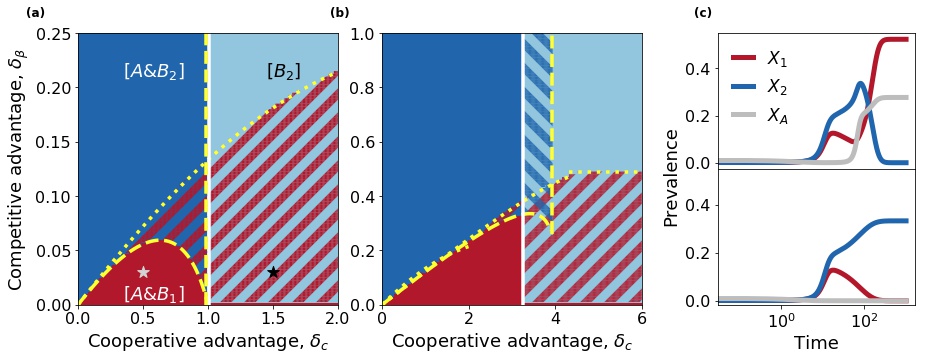}
\caption{\textbf{Phase diagram for the well-mixed system}. (a),(b) Stable equilibria as a function of $\delta_{\beta}$ and $\delta_{c}$ for two parameter choices, namely (a) $\alpha = 0.6$, $\beta_2 = 1.5$, $c_1 = 4$, and (b) $\alpha = 0.8$, $\beta_2 = 1.1$, $c_1 = 7$. The three states $[B_2]$, $[A\&B_2]$ and $[A\&B_1]$ are indicated in light blue, dark blue and red, respectively. Hatched regions correspond to bistable and multistable regions. The yellow curves show the analytical boundaries delimiting stability regions for $[A\&B_1]$ and $[A\&B_2]$ (equations~(\ref{eq:Abi}) and (\ref{eq:stab_ABi})), while the white one delimits the $[B_2]$'s region (equation~(\ref{eq:bB2})). Notice that for $\delta_{c}>3, \, 6$, for panels (a) and (b) respectively, $c_2<1$ and the interaction between $B_2$ and $A$ ceases to be cooperative. This provides naturally a range for x-axis. In panel (b) $\beta_1$ is below one for $\delta_\beta > 0.1$. 
(c) Evolution of total prevalence for $A$ (grey), $B_1$ (red) and $B_2$ (blue), considering singly and doubly infected combined. Parameters correspond to the grey and black star markers in panel (a), i.e. $\delta_\beta = 0.03$ and $\delta_c = 0.5, \, 1.5$ in top and bottom panels, respectively. Dynamical trajectories have been obtained by integrating equations~\eqref{Eq:system} with initial conditions: $B_i(t=0)=0.001$, $A(t=0)=0.01$.
}
\label{fig:Fig2}
\end{figure}

Figures~\ref{fig:Fig2}a,b show the location of stable states with two combinations of $\alpha$, $\beta_2$ and $c_1$. Results that are obtained for other parameter values are reported in figure~S1. No co-existence was found between $B_1$ and $B_2$. In principle, equations~\eqref{Eq:system} admit a co-existence equilibrium $[A \& B_1 \& B_2]$. However, this co-existence was always found to be unstable in the numerical simulations. Persistence of $A$ is only possible together with one of the $B$ strains. The equilibrium solution $[A]$ is unfeasible for $\alpha<1$ and unstable for $\alpha>1$, unless both reproductive ratios, $\beta_i$, are below the epidemic threshold.
Because of the assumption $\delta_{\beta} > 0$, $B_2$ outcompetes $B_1$ in absence of $A$, in agreement with the principle of competitive exclusion. Therefore the final state $[B_1]$ is always unstable, and persistence of $B_1$ is possible only in co-circulation with $A$. On the other hand, $B_2$ can spread either alone or together with $A$. Specifically, the $[B_2]$ configuration is feasible for $\beta_2 > 1$. It is stable if and only if $\alpha<\alpha_c$, with
\begin{equation}
\alpha_c= \dfrac{\beta_2}{c_2 (\beta_2 - 1) + 1} \: .
\label{eq:bB2}
\end{equation}
\noindent This provides a sufficient condition for the persistence of $A$. Equation~\eqref{eq:bB2} can be expressed in terms of $\delta_c$, namely $\delta_c > c_1 - (\beta_2-\alpha)/[\alpha(\beta_2-1)]$, which is visualized as the white boundary in figure~\ref{fig:Fig2}a,b.

The competition between $B_1$ and $B_2$ is governed by the trade-off between transmission and cooperative advantage. This is described by the boundaries of the $[A \& B_i]$ regions that can be traced by combining the feasibility and stability conditions. These boundaries are plotted in figures~\ref{fig:Fig2}a,b as dotted and dashed yellow curves for $[A\&B_1]$ and $[A\&B_2]$, respectively. 
For a solution to be feasible the densities of all states must be non-negative. For absolute parameter values as in figure~\ref{fig:Fig2}a,b we found that this yields the necessary condition $\alpha \beta_i > 4(c_i -1)/c_i^2$, corresponding to the vertical and horizontal segments. On the other hand, the stability boundary separating $[A \& B_i]$ from any state containing $B_j$ ($j \neq i$) is given by 
\begin{equation}
\beta_j (S^* + c_j A^*) - 1 < 0,   
\label{eq:Abi}
\end{equation}
\noindent where $S^*$ and $A^*$ are the equilibrium densities of $S$ and $A$, respectively, evaluated in the configuration $[A \& B_i]$.
The left-hand side of the equation represents the growth rate of the competitor $B_j$, appearing in equation~\eqref{Eq:Xi}, and evaluated in the $[A \& B_i]$ state. Thus, the relation~\eqref{eq:Abi} expresses the condition for $B_j$ extinction.
Expressed in terms of $\delta_c$ and $\delta_{\beta}$, the conditions becomes:
\begin{eqnarray}
    \left[A \& B_1\right] &:& \dfrac{\beta_2 (c_1-\delta_c)}{ c_1(\beta_2-\delta_\beta)}  + \frac{\beta_2\delta_c}{c_1-1}  \left(1 - \sqrt{1-\dfrac{4(c_1-1)}{c_1^2(\beta_2-\delta_\beta)\alpha} } \right)=1 
    \\ 
   \left[A \& B_2\right] &:& \dfrac{c_1(\beta_2-\delta_\beta)} {\beta_2 (c_1-\delta_c)} - \dfrac{\delta_c(\beta_2-\delta_\beta)}{c_1-\delta_c-1} \left(1 - \sqrt{1-\dfrac{4(c_1-\delta_c-1)}{\beta_2\alpha(c_1-\delta_c)^2}} \right)=1. \nonumber
        \label{eq:stab_ABi}
\end{eqnarray}

The intersection among the stability boundaries described above produces a rich state space.
For all tested values of $\alpha$, $\beta_2$ and $c_1$, we found a wide region of the $(\delta_c, \delta_\beta)$ space (red-hatched in figures~\ref{fig:Fig2}a,b) displaying bistability between the $[A\&B_1]$ state and a $B_2$-dominant state with either $[B_2]$ or $[A\&B_2]$. 
In certain cases, bistability can also occur between the $[B_2]$ and $[A\&B_2]$ states (blue-hatched region in figure~\ref{fig:Fig2}b). 
This has been studied in the past for two cooperating pathogens~\cite{chen_fundamental_2017}. 
We found that the intersection between the latter region and the red-hatched region gives rise to a multistable state. 

Interestingly, for all tested parameters we found that the boundary of the $[A\&B_2]$ stability region is not monotonic. As a consequence, for a fixed $\delta_{\beta}$ a first transition from the $[A\&B_2]$ state to $[A\&B_1]$ is found for small $\delta_{c}$ values. An increase of $\delta_{c}$ leads to a second boundary with a bistable region, where the dominance of $B_1$ over $B_2$ depends on initial conditions. The transition for small $\delta_c$ is expected: By increasing $B_1$'s advantage in cooperation, a point is reached beyond which $B_1$'s disadvantage in transmissibility is overcome. On the other hand, the second threshold appears to be counter-intuitive. We investigated it more in depth for the case depicted in figure~\ref{fig:Fig2}a. We plotted the infectious population curves as a function of time for each infectious compartment. We compared $\delta_c= 0.5$, which corresponds to the $[A\&B_1]$ stable state (figure~\ref{fig:Fig2}c top), and $\delta_c= 1.5$, which leads to a bistable region (figure~\ref{fig:Fig2}c bottom), where all other parameters are as in figure~\ref{fig:Fig2}a. Figure~\ref{fig:Fig2}c shows that $B_1$ loses the competition at the beginning. However, when $B_2$ is sufficiently cooperative with $A$ (top), the rise of $B_2$ leads to a rise in $A$ that ultimately drives $B_1$ to dominate. For higher $\delta_c$ the strength of cooperation between $B_2$ and $A$ is not sufficient. The indirect beneficial effect of $B_2$ over $B_1$ is not present (bottom), and $B_1$ can dominate only if initial conditions are favourable. 

\begin{figure}[ht!]
\centering
\includegraphics[width=\textwidth,keepaspectratio]{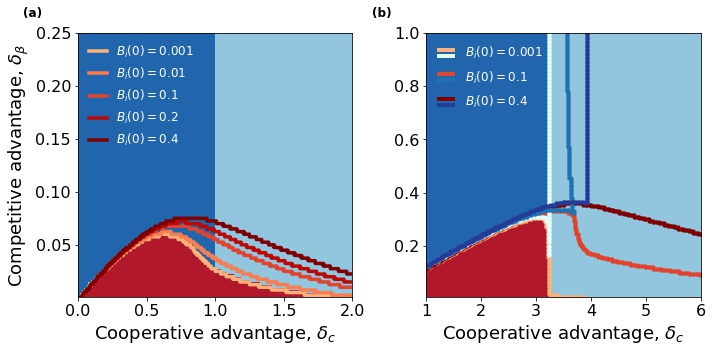}
\caption{\textbf{Equilibrium configurations for the well-mixed system}. Final outcome obtained by numerically integrating equations~\eqref{Eq:system} for $B_i(t=0)=0.001$, $A(t=0) = 0.01$. (a) $\alpha = 0.6$, $\beta_2 = 1.5$, $c_1 = 4$. Boundaries of the $[A\&B_1]$ state for different initial conditions are indicated by red-scale contours. (b) $\alpha = 0.8$, $\beta_2 = 1.1$, $c_1 = 7$. Here, the boundaries of the $[A\&B_2]$ are shown (in blue shades), together with the ones of $[A\&B_1]$. }
\label{fig:Fig3}
\end{figure}

In the bistable and multi-stable regions, the outcome of the competition is determined by initial conditions. While a mathematical analysis is complicated due to the multi-dimensionality of the problem, we gained insights into the basins of attraction by numerically integrating equations~\eqref{Eq:system} while exploring different combinations of $B_i(t=0)$ and $A(t=0)$. For the bistability between the regions of $B_1$ and $B_2$ dominance, we considered the parameter combination of figure~\ref{fig:Fig2}a and show in figure~\ref{fig:Fig3}a the states that are reached starting from $B_i(t=0) = 0.001$ and $A(t=0) = 0.01$. The bundle of curves with different shades of red (from light to dark) indicates the boundary of the $[A\&B_1]$ equilibrium when $B_1(t=0)$ and $B_2(t=0)$ are equally increased. We found that an increase in $B_1$'s initial infected densities favours the $[A\&B_1]$ state, as expected. Interestingly, however, an increase in $B_1(t=0)$ results in the $[A\&B_1]$ region to expand even when $B_2$'s density increases at the same level. Figure~\ref{fig:Fig3}b shows that a similar behaviour is found when parameters are as in figure~\ref{fig:Fig2}b. In this case, the 
region $[A \& B_2]$ expands together with the $[A \& B_1]$ one. Thus, increased initial frequencies promote co-circulation between $B$ and $A$.
In figure~S2 we present a deeper exploration of initial conditions, considering the parameter combination of figure~\ref{fig:Fig2}a as an example. We found that an increase in the initial level of $A$ also favours $B_1$. However, the initial advantage (either in $B_1(0)$ or $A(0)$) that is necessary for $B_1$ to win against $B_2$ increases as $\delta_\beta$ increases.

The stability diagrams obtained with several parameter sets, explored in a latin-square fashion, is reported in figure~S1. This shows that increased transmissibility and cooperativity levels enhance the cooperative interaction of $B_i$ strains with $A$. This results in an increase in the parameter region for which $B_1$ together with $A$ dominates over $B_2$. For instance, the comparison between panels (d) and (f) in the figure shows that, by increasing $\beta_2$ from 1.1 to 1.5, the same difference in strain epidemiological traits, $\delta_c$ and $\delta_\beta$, may lead  to a switch in dominance from $B_2$ to $B_1$.

\subsection{Continuous system with communities}
\label{sec:continuous-communities}

\begin{figure}[ht!]
\centering
\includegraphics[width=\textwidth,keepaspectratio]{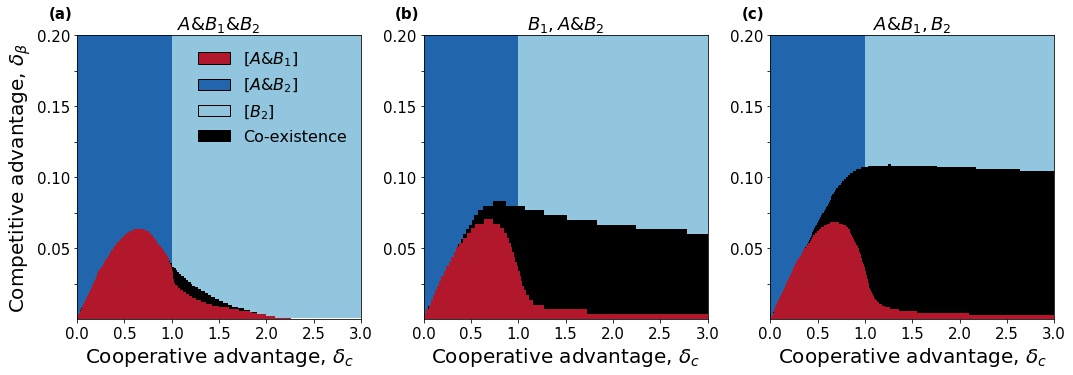}
\caption{\textbf{Equilibrium configurations for two interacting communities}. Final outcome obtained by numerically integrating the equations when: (a) all strains start in the same community (together with $A$); (b) $B_1$ and $B_2$ start in separate communities, with $A$ starting together with $B_2$; (c) $A$ starts along with $B_1$, while $B_2$ starts separately. Initial density of each pathogen/strain is 0.01. Here $\epsilon = 0.0002$. Other parameters are as in figure~\ref{fig:Fig2}a. 
}
\label{fig:Fig4}
\end{figure}

\begin{figure}[ht!]
\centering
\includegraphics[width=\textwidth,keepaspectratio]{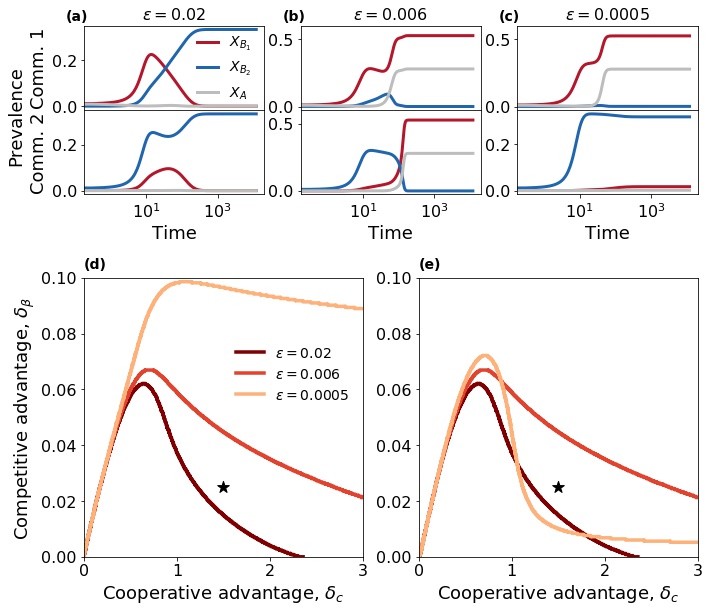}
\caption{\textbf{Role of spatial separation for two interacting communities}.
(a-c) Dynamical trajectories within community 1 and 2 obtained for (a) $\epsilon = 0.02$, (b) $\epsilon = 0.006$, and (c) $\epsilon = 0.0005$. (d,e) Boundaries in the $\delta_c$, $\delta_{\beta}$ plane delimiting the regions where the dynamics ends up in: (d) $B_1$ persistence (i.e. $[A \& B_1]$ or full co-existence); (e) $[A \& B_1]$ state. In all panels, $A$ and $B_1$ are seeded together into one community, while $B_2$ is seeded into the other community; the initial density of each species is set to 0.01. Trajectories are obtained by setting $\delta_c = 1.5$, $\delta_{\beta} = 0.025$ (black star in panels (d),(e)). Other parameters are as in figure~\ref{fig:Fig2}a.}
\label{fig:Fig5}
\end{figure}

We now consider a population that is divided into two communities (cf. figure~\ref{fig:Fig1}c).  For simplicity, we assumed that they are of the same size. To differentiate transmission within and across communities, we rescaled the force of infection produced by individuals of a different community by a factor $\epsilon$, and the force of infection of individuals of the same community by $1-\epsilon$. We assumed $0 < \epsilon \leq \tfrac{1}{2}$ in order to consider the case in which individuals mix more within their community than outside - the limit $\epsilon=\frac{1}{2}$ corresponds to homogeneous mixing. 

Given the high number of variables, a stability analysis is difficult in this case. Still, the dynamics can be reconstructed through numerical integration of the equations. Figure~\ref{fig:Fig4} shows the final states with fixed $\epsilon$, $\beta_2$, $C_1$ and $\alpha$. Other parameter values are analysed in figure~S3. 
Figures~\ref{fig:Fig4}a-c compare different seeding configurations, while keeping the initial density of each pathogen/strain to 0.01: (a) all strains are seeded in community 1 and community 2 is completely susceptible; (b) $B_1$ is seeded in community 1 while $B_2$ and $A$ are initially present in community 2 only; (c) $B_1$ and $A$ are seeded together in community 1, while $B_2$ is seeded in community 2. 
In all cases we found a diagram with shape similar to figure~\ref{fig:Fig3}a. However, a new region is now present (indicated in black) where all players co-exist. This occurs when strains are separated since the beginning -- see figure~S4 for additional seeding configurations. Interestingly, however, this happens also for a tiny region of the parameter space, when all strains are seeded together (panel a), provided that the other community is initially disease-free. 

Figure~\ref{fig:Fig5} sheds light on the dynamics leading to the outcomes of figure~\ref{fig:Fig4}. In order to benefit from the cooperative advantage, the $B_1$ incidence must be above a certain threshold. Figures~\ref{fig:Fig5}b,c show that incidence of $A$ remains close to zero, until incidence of $B_1$ is sufficiently high. 
With $B_2$ seeded on a different community (community 2), the direct interaction between the two strains is delayed by the time necessary for $B_2$ to reach the community of $B_1$. For high $\epsilon$ the delay is short and $B_2$ reaches community 1 before $A$ incidence starts to raise (figure~\ref{fig:Fig5}a). On the other hand, for lower $\epsilon$, $B_1$ has enough time to build up a cooperative protection before the arrival of $B_2$. This makes it resistant to the invader. At intermediate $\epsilon$, $B_1$ becomes able to overcome $B_2$ in community 2. For small $\epsilon$, strains spread in their origin community independently from one another.

In summary, a decrease in $\epsilon$ increases the region of $B_1$ persistence (figure~\ref{fig:Fig5}d). However, this may be associated to either $B_1$ dominance or co-existence. Reducing the values of $\epsilon$, the region corresponding to the $[A\&B_1]$ state expands first and shrinks later, leaving the place to the co-existence region. This is shown by the non-monotonous change of the $[A\&B_1]$ region in figure~\ref{fig:Fig5}e.

When all strains start in the same community, co-existence is enabled by a segregation mechanism similar to the one described above. In this case, separation occurs during the early stage: $B_2$ rapidly spreads in the other community due to its advantage in transmissibility, and becomes dominant there (cf. figure~S5). This enables co-existence in a parameter region where $B_1$ would otherwise dominate.

Results described so far were obtained with fixed values of $\beta_2$, $C_1$ and $\alpha$. Additional parameter choices are shown in figure~S3. Increasing in $\alpha$ was found to enlarge the $B_1$ dominance region, as in the well-mixed case. In addition, co-existence becomes possible for $\alpha>1$ in a very small region of the parameter space.

\subsection{Spreading on networks}
\label{sec:results_networks}
The continuous deterministic framework analysed so far does not account for stochasticity and for the discrete nature of individuals and their interactions. These aspects may alter the phase diagram and shape the transitions across various regions. We casted our model on a discrete framework in which individuals are represented by nodes in a static network. Possible individual states are still the same as in the mean-field formulation, and infection can spread only between neighbouring nodes. We first considered an Erd\H{o}s-Rényi graph, where the mixing is homogeneous across nodes. Denoting $N$ the number of nodes and $\bar k$ the average degree, the network was built by connecting any two nodes with probability $\bar k /(N-1)$. We run stochastic simulations of the dynamics. In order to see the effect of multi-pathogen interactions, we minimised the chance of initial stochastic extinction by infecting a relatively high number of nodes at the beginning: 100 infected for each infectious agent. We then computed the fraction of stochastic simulations ending up in any final state, the average prevalence for each strain ($X_1$, $X_2$) in the final state and the average coexistence time. Additional details on the network model and the simulations are reported in the Supplementary Material.

The phase diagram of figure~\ref{fig:Fig6}a is similar in many aspects to its continuous deterministic version (figure~\ref{fig:Fig3}a). Three final states are possible, i.e. $[B_2]$, $[A\&B_1]$ and $[A\&B_2]$ (figure~\ref{fig:Fig6}a). Here, however, the same initial conditions and parameter values can lead to different stochastic trajectories and stationary states. For instance, the red region in the figure corresponds to the case in which the final state $[A\&B_1]$ is reached very frequently; however, the dynamic trajectories can end up also in the $[A\&B_2]$ or in the $[B_2]$ states. The transitions across the different regions of the diagrams can be very different, as demonstrated by figures~\ref{fig:Fig6}b-j. Panels b,c,d show the effect of varying $\delta_{c}$ at a fixed $\delta_{\beta}$. The transition between $[A\&B_2]$ and $[A\&B_1]$ on the left is sharp. Both the probability of one strain winning over the other and the equilibrium prevalence change abruptly for a critical value of $\delta_{c}$. Here, the spreading is super-critical for all pathogens: $\beta_1, \beta_2>1$ and $c_1$, $c_2$ are sufficiently high to sustain the spread of $A$. The transition is due to the trade-off between $B_1$ and $B_2$ growth rates. Conversely, the probability of ending up in the $[B_2]$ state rises slowly, driving the gradual transition from the red to the light blue region on the right. This region appears in correspondence of the bistable region of the continuous/deterministic diagram -- figure~\ref{fig:Fig2}. Here, $A$ undergoes a transition from persistence to extinction, driven by the drop in $c_2$ (figure~S6). This critical regime is characterised by enhanced stochastic fluctuations.
When $\delta_{c}$ is fixed and $\delta_{\beta}$ varies, we found a sharp transition (panels e,f,g) and a hybrid transition, where the final state probability varies gradually and the equilibrium prevalence ($X_1$) varies abruptly (panels h,i,j). 

\begin{figure}[htbp]
\centering
\includegraphics[width=0.7\textwidth]{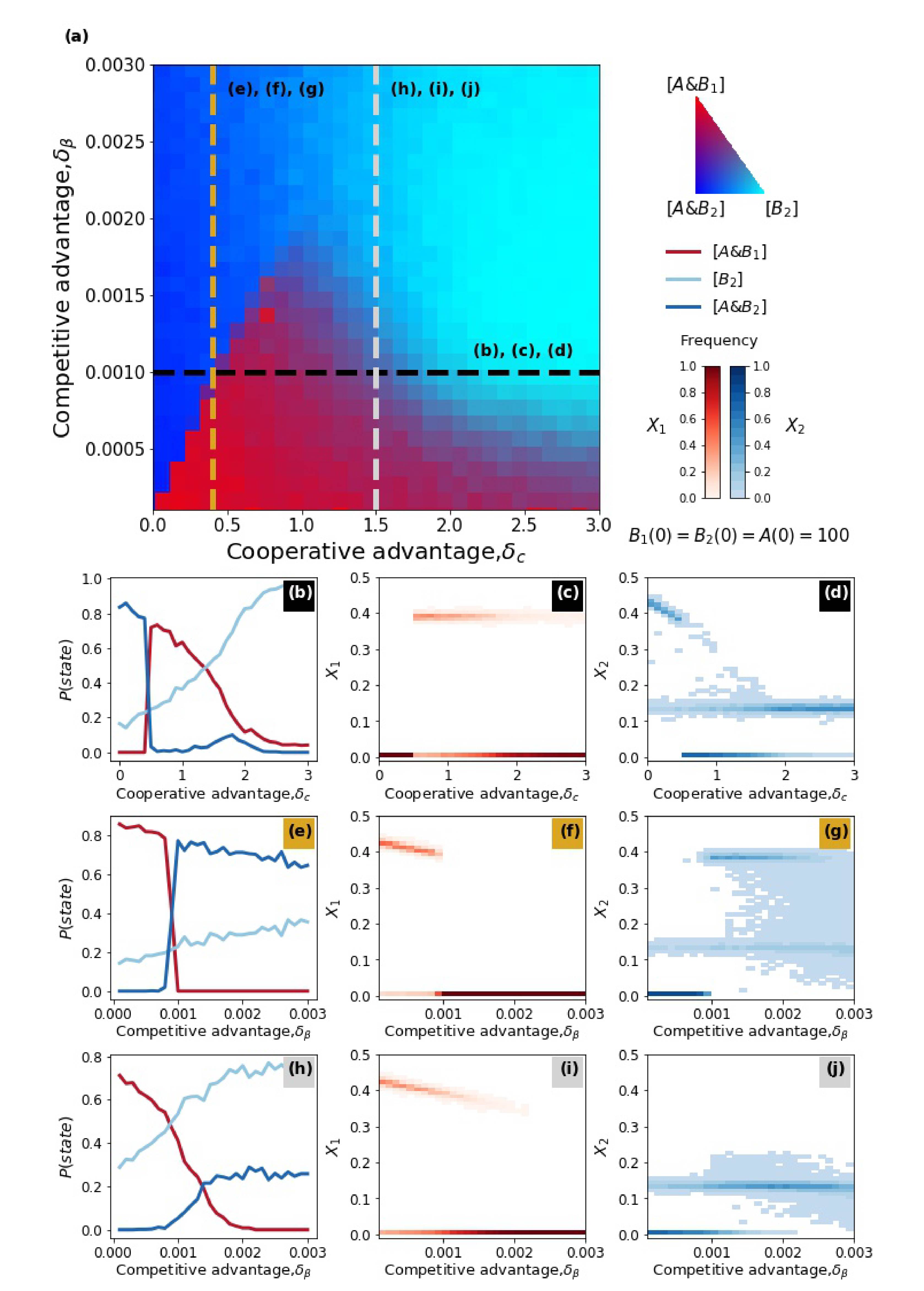}
\caption{\textbf{Phase diagram for the Erd\H{o}s-Rényi network}. (a) Frequency of stationary states, as obtained in numerical simulations. The colour scale in the legend quantifies the proportion of runs ending in the different states among $[B_2]$, $[A\&B_2]$ and $[A\&B_1]$. Here, the extremes of the colour map correspond to the case in which these states are found in 100\% of runs. Initial conditions are shown in the figure.  (b-j) Equilibrium state probability (left column), and distribution of both $B_1$'s and $B_2$'s prevalence in the final state (middle and right columns respectively) along the dashed lines in panel (a). Specifically: $\delta_{\beta} = 0.001$ for (b), (c) and (d); $\delta_{c} = 0.4$ for (e), (f) and (g); $\delta_{c} = 1.5$ for (h), (i) and (j). For convenience, we reparametrised the model taking the time step as unit of time: $\Delta t = 1$. We set the following parameters' values: $\mu = 0.05$, $\alpha = 0.009$, $\beta_2 = 0.015$, $c_1 = 4$, $N=20000$, $\bar k = 4$. Note that in the case of spread on networks we have $R^{(i)}_0=\beta_i \rho/\mu$ and $R^{(A)}_0=\alpha \rho/\mu$, where $\rho$ is the spectral radius of the adjacency matrix. For the $\alpha$, $\beta_2$ and $\delta_{\beta}$ values considered in the figure we have $R^{(1)}_0, R^{(2)}_0>1$.}
\label{fig:Fig6}
\end{figure}

We concluded by analysing the effect of community structure. 
Each node was assigned to one among $n_C$ communities, which we assumed for simplicity to have equal size $N/n_C$, and has a number of open connections drawn from a Poisson distribution with average $\bar k$.
Links were formed by matching these connections according to an extended configuration model, where a fraction $\epsilon$ of stubs connects nodes of different communities. In this way the model is the discrete version of the one in Section~\ref{sec:continuous-communities}.

\begin{figure}[htbp]
\centering
\includegraphics[width=\textwidth]{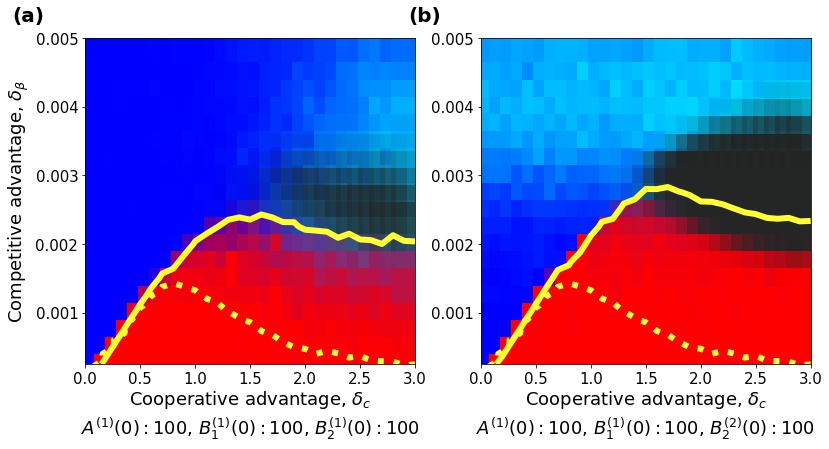}
\caption{\textbf{Phase diagram for the random modular network}. Frequency of equilibrium configurations, as obtained in the numerical simulations, with (a) $A$, $B_1$ and $B_2$ starting in the same community, and (b) $A$ and $B_1$ starting from a different community to $B_2$. Detailed initial conditions are directly shown on each panel. For each player, the superscript $i$ indicates the community where the infectious are seeded. The colour scale is the same as in figure~\ref{fig:Fig6}. The frequency of runs for which co-existence of all strains was observed after $T_{max}=2\cdot 10^{6}$ time steps is shown with different shades of black. Contour lines representing the 0.5 probability to end up in the $[A\&B_1]$ state are indicated to enable a comparison between the Erd\H{o}s-Rényi (dashed line) and the random modular network (continuous line). We considered $n_C = 10$ and $\epsilon = 0.003$. Other parameter values are as in figure~\ref{fig:Fig6}.}
\label{fig:Fig7}
\end{figure}

Mean-field results remain overall valid. The two plots in figure~\ref{fig:Fig7} mirror panels a,c of figure~\ref{fig:Fig4} and show a similar behaviour. We find evidence of a co-existence region (in black in the figure), where no extinction is observed during the simulation time frame - here set to $2\cdot 10^6$ time steps, around two orders of magnitude longer than the time needed to observe strain extinction in the Erd\H{o}s-Rényi case. Such region is larger when the two strains are seeded in separated communities (figure~\ref{fig:Fig7}b), but it is still visible when strains start altogether (figure~\ref{fig:Fig7}a). Co-existence occurs less frequently in the latter case, since it requires strains to reach the separation during the spreading dynamics.

Analogously to the continuous deterministic model we found that the separation in communities favours the more cooperative strain. The region where $B_1$ wins is larger compared to the Erd\H{o}s-Rényi case (as highlighted by the comparison between the dashed and the continuous curves). In addition, the probability of winning is close to one for a large portion of the $[A\&B_1]$ dominance region. 

\section{Discussion and conclusion}
\label{sec:discussion}
We presented here a theoretical analysis of a three-player system where both competition and cooperation act simultaneously. We have considered two competing strains co-circulating in the presence of another pathogen cooperating with both of them. Strains differ in epidemiological traits, with one strain being more transmissible but less cooperative than its competitor. Through mathematical analyses and computer simulations we have reconstructed the possible dynamical regimes, quantifying the conditions for dominance of one strain or co-existence. We found that the interplay between competition and cooperation leads to a complex phase diagram whose properties cannot be easily anticipated from previous works that considered competition and cooperation separately. 

We showed that it is possible for a more cooperative strain to dominate over a more transmissible one, provided that the difference in transmissibility is not too high. This suggests that the presence of another pathogen ($A$) might alter the spreading conditions, creating a favourable environment for a strain that would be otherwise less fit. While dominance depends on the difference in epidemiological traits, we found that variations in the absolute cooperation and transmissibility levels may change the hierarchy between strain -- analogously to~\cite{gjini_slow-fast_2017} -- with a higher spreading potential of either $B_i$ or $A$ favouring the more cooperative strain. 

Interestingly, the cooperative strain can dominate also when $A$ has a sub-critical reproductive ratio ($\alpha<1$) -- when spreading alone -- and relies on the synergistic interaction with $B$ strains to persist. 
The dynamical mechanisms underlying this outcome are complex. We analysed a case with a small difference in cooperativity, and we found that the more transmissible strain, by spreading initially faster, creates the bulk of $A$ infections that in turn favour its competitor.
In other words, direct competition for susceptible hosts is not the only force acting between strains: an indirect, beneficial interaction is also at play, mediated by the other pathogen. 
The dominance outcome is thus the result of the trade-off between these two forces. When the difference in cooperation is higher, two or more stationary configurations are possible.
In this scenario, the final outcome is determined also by the initial frequency of each pathogen/strain. 
We found that, in certain situations, an initial advantage of one strain is able to drive it to dominance. This is in contrast with simpler models of competition, where the final outcome is determined solely by the epidemiological traits.
The outcome, however, is also governed by pathogen $A$ that favours the more cooperative strain.
Previous works have analysed multistability in two-pathogen models with cooperation in relation to the hysteresis phenomenon, where the eradication threshold is lower than the epidemic one~\cite{chen_fundamental_2017,cai_avalanche_2015,cui_mutually_2017}. A similar mechanism could be at play here. However, the identification of hysteresis loops requires a better reconstruction of the attraction basins. While the numerical work presented here provided some preliminary understanding, a deeper mathematical analysis would be needed in this direction. Multistability is, instead, not present in two-pathogen models with complete mutual exclusion. This dynamical feature emerges, however, in the more general case where strains are allowed to interact upon co-infection~\cite{gjini_slow-fast_2017}.

While we did not find stable co-existence among strains in the well-mixed system, co-existence was possible in presence of community structure. In this case, strains can minimise competition for hosts through segregation. Importantly, spatial separation alone is not sufficient to enable co-existence between two strains, when complete mutual exclusion is assumed. This was already known from previous works that showed that community structure must be combined with some level of heterogeneity across communities to enable co-existence, e.g. a strain-specific adaptation to a population or environment to create an ecological niche~\cite{wu_coexistence_2016,allen_sis_2004,feng_coexistence_2011,qiu_vectorhost_2013,lin_global_2014}. Here, communities are homogeneous and co-existence is the result of the interplay between community structure and presence of the cooperative pathogen. When the two strains are seeded in different communities, their interaction occurs after the time lag necessary for one strain to invade the other community. We found that this interval may allow the resident strain to reach the bulk of infections necessary to fend off the invasion. This mechanism is rooted again in the effect of pathogens' frequencies on strain selective advantage. The drivers of strains' co-existence remain an important problem in disease ecology with applications to both vaccination and emergence of anti-microbial resistance. Within-host and population factors have been studied in the past by several modelling investigations. 
Notably, while co-existence is not possible in models with complete mutual exclusion, this may be enabled in co-infection models~\cite{gjini_slow-fast_2017,gjini_how_2016,sofonea_within-host_2015,alizon_co-infection_2013,alizon_multiple_2013}. Other models have addressed environmental and host population features, such as age-structure, contacts dynamics and spatial organisation~\cite{martcheva_subthreshold_2007,cobey_host_2017,eames_coexistence_2006,buckee_host_2007}. However, little attention has been dedicated to the effect of an additional co-circulating pathogen. Cobey et al. studied the interaction between \textit{Haemophilus influenzae} and \textit{S. pneumoniae} co-circulating strains~\cite{cobey_pathogen_2013}. Despite the numerous differences between our model and theirs, their work provides results consistent with ours. Namely, the multi-strain dynamics can be affected by another pathogen.

We simulated the three-player dynamics on networks and we obtained phase diagrams that are similar to the continuous-deterministic counterparts. The discrete/stochastic framework, however, allows for observing the nature of the phase transitions. Several works recently studied the nature of the epidemic transition for two cooperating pathogens, highlighting differences with the single-pathogen case. Cooperation was found to cause discontinuous transitions where the probability of an outbreak and prevalence change abruptly around a critical value of the transmission rate~\cite{hebert-dufresne_complex_2015, chen_fundamental_2017}, akin to other complex contagion mechanisms such as the ones found in social contagion~\cite{gomez-gardenes_explosive_2016,dodds_universal_2004}. This phenomenon, however, is sensitive to the network topology, with continuous, discontinuous and hybrid, i.e. continuous in the outbreak probability and discontinuous in the prevalence, transitions observed according to the topology of the network~\cite{hebert-dufresne_complex_2015,chen_fundamental_2017,cai_avalanche_2015,cui_mutually_2017,wang_coevolution_2019,rodriguez_particle_2019, choi_two_2018}. Here we found rich dynamics as the impact of stochastic effects. These effects were less important when the difference between strains' epidemiological traits was small. Conversely, for a higher difference in cooperative factor different outcomes are equally probable. 
Results presented here are preliminary and limited to two network configurations. Future work should investigate additional network topologies, e.g. a power-law degree distribution, and further values of the network parameters. In addition, more sophisticated numerical analysis (e.g. scaling analysis) would be needed to better classify the nature of the phase transitions.

Concurrence of inter-species cooperation and intra-specie competition is present in many epidemiological situations. 
Currently, around 90 distinct \textit{S. pneumoniae} serotypes are known to co-circulate worldwide, despite indirect competition mediated by host immune response~\cite{cobey_niche_2012}. The emergence of antibiotic-resistant strains and the development of vaccines able to target only a subset of strains has motivated extensive research on the drivers of \textit{S. pneumoniae} ecology~\cite{cobey_niche_2012,cobey_host_2017,opatowski_assessing_2013}. Strain circulation is facilitated by respiratory infections, e.g. influenza~\cite{shrestha_identifying_2013, opatowski_influenza_2018} and some bacterial infections~\cite{bosch_viral_2013,dunne_nasopharyngeal_2013}.
Cooperative behaviour has been observed also between HIV and infections such as HPV, tuberculosis and malaria~\cite{looker_evidence_2018,cohen_increased_2005, abu-raddad_dual_2006,tornheim_tuberculosis_2017,wells_hiv_2007}. This increases the burden of these pathogens and causes public health concern. At the same time, there is evidence that different strains of tuberculosis~\cite{cohen_modeling_2008,dowdy_data_2013}, malaria~\cite{chen_competition_2014}, and HPV~\cite{pons-salort_exploring_2013,gray_occurrence_2019,murall_revising_2014} may compete. In particular, multidrug-resistant strains of tuberculosis (MDR-TB) are widely spread, although the acquisition of resistance seems to be associated to a fitness cost~\cite{wells_hiv_2007,gagneux_fitness_2009}. The synergistic interaction with HIV could play a role in this emergence and surveillance data suggest a possible convergence between HIV and MDR-TB epidemics in several countries~\cite{wells_hiv_2007}. Our theoretical work highlights ecological mechanisms potentially relevant to these examples.
In this regard, an essential aspect of our model is the trade-off between transmissibility and cooperativity in determining strain advantage. Although differences in transmissibility across strains have been documented, e.g. fitness cost of resistance~\cite{melnyk_fitness_2015}, gathering information on strain-specific cooperative advantage remains difficult. The theoretical results illustrated here show the importance of quantifying this component for better describing pathogen ecosystems.

This study also represents the starting point of more complex models where multiple strains are involved and competition and cooperation are acting simultaneously. Patterns of competitive and cooperative interactions could be at play for instance among recently emerged pathogens such as Zika virus~\cite{rothan_current_2018}. Zika virus has emerged in regions where Dengue and Chikungunya viruses are endemic. Observed patterns of sequential monodominance by one arbovirus at a time at a given location suggest competition between these pathogens~\cite{magalhaes_zika_2017}. Also, considerable effort is currently devoted to characterising possible positive interactions between Zika virus and HIV~\cite{rothan_current_2018}. In some cases, different strains of the same pathogen can interact both competitively and cooperatively, as in the case of Dengue~\cite{reich_interactions_2013,shrestha_statistical_2011}. Primary Dengue infections are characterised by mild symptoms and grant short-term cross-protection against other serotypes. As cross-immunity wanes over time, however, secondary Dengue infections not only become possible but are also associated with severe illness and with increased virulence. 

The examples above involve diseases with varying natural history and time scales and should be modelled with different compartmental models -- SIR, SI, SIS, SIRS. We decided here to consider two SIS pathogens and the results cannot be readily extended to other models, since the dynamics of disease unfolding alters the outcome of strain interactions. It is important to notice, however, that several dynamical properties of competitive and cooperative interactions, such as dominance vs. co-existence~\cite{karrer_competing_2011} and abrupt transitions~\cite{grassberger_phase_2016,chen_outbreaks_2013,zarei_exact_2019}, hold for both SIS and SIR. 

The model studied here is based on certain simplifications. All pathogens are assumed to have the same recovery rate; moreover, cooperation acts in both directions and the same factors $c_i$ quantify the enhancement in susceptibility when $A$ infection occurs before $B_i$ infection and vice versa. These assumptions may not hold for many synergistic pathogens, especially when cooperative benefits are based on different biological mechanisms. For instance, while HIV increases susceptibility against \textit{P. falciparum}, the latter increases HIV's viral load, thus increasing HIV's virulence rather than host susceptibility to HIV~\cite{cohen_increased_2005, abu-raddad_dual_2006}. It is likely that, by relaxing these assumptions, our model could exhibit even more complex phase diagrams. Eventually, other aspects of the disease-specific mechanisms and multi-pathogen interactions could affect the results presented here and should be addressed in future works. These include latent infections, which are characteristic, for instance, of tuberculosis~\cite{cohen_modeling_2008}, partial mutual exclusion among strains~\cite{cohen_modeling_2008,obadia_detailed_2015, gjini_slow-fast_2017,sofonea_within-host_2015}, or interaction mechanisms other than the ones introduced here (e.g. affecting the infectious period~\cite{sanz_dynamics_2014}). 

In conclusion, we have provided a theoretical study of a dynamical system where both competition and cooperation are at play. We found that a less transmissible and more cooperative strain may dominate; however, the conditions on the parameters for this to happen are non-trivial (non-monotonic) and the outcome critically depends on initial conditions and stochastic effects. When coupled with population structure, the presence of a cooperative pathogen may create the conditions for multi-strain co-existence by dynamically breaking the spatial symmetry and creating ecological niches. These results provide novel ecological insights and suggest mechanisms that may potentially affect the dynamics of interacting epidemics that are of public health concern.

\section*{Data availability}
The python code used for the mean-field analyses and the C++ code for the stochastic simulations on networks are publicly available at the following link:

\noindent https://github.com/francescopinotti92/Competition-Cooperation.

\section*{Authors' contributions}
FP, FGh, PH, CP conceived and designed the study. FP carried out the analyses. FP, FGh, PH, CP analysed the results. FP, FGh, PH, CP contributed to the writing of the manuscript and approved the final version.

\section*{Competing interests}
The authors declare no competing interests.

\section*{Funding}
FP acknowledges support by Pierre Louis Doctoral School of Public Health. FGh and PH acknowledge support by German Academic Exchange Service (DAAD) within the PPP-PROCOPE framework (grand no. 57389088). CP and FP acknowledge support by the Partenariats Hubert Curien within the PPP-PROCOPE (grand no. 35473TK). FGh acknowledges partial support by Deutsche Forschungsgemeinschaft (DFG) under the grant GH 176/1-1 (idonate project: 345463468). PH acknowledges further support by DFG in the framework of Collaborative Research Center 910 (grand no. SFB910). CP acknowledges support from the municipality of Paris through the program Emergence(s). 

\section*{Research ethics}
This section does not apply to this study.

\section*{Animal ethics}
This section does not apply to this study.

\section*{Permission to carry out fieldwork}
This section does not apply to this study.

\section*{Acknowledgements}
We thank Andrew Keane for helpful discussions that made the manuscript more accessible.

\bibliographystyle{unsrt}



\section*{Supplementary material}

# Supplementary material for: Interplay between competitive and cooperative interactions in a three-player pathogen system

Francesco Pinotti[1], Fakhteh Ghanbarnejad[2,3,4], Philipp Hövel[2,5], and Chiara Poletto[1,*]

[1]INSERM, Sorbonne Université, Institut Pierre Louis d'Épidémiologie et de Santé Publique, IPLESP, F75012, Paris, France
[2]Institut für Theoretische Physik, Technische Universität Berlin, Hardenbergstraße 36, 10623 Berlin, Germany
[3]The Abdus Salam International Centre for Theoretical Physics (ICTP), Trieste, Italy
[4]Physics Department, Sharif University of Technology, P.O. Box 11165-9161, Tehran, Iran
[5]School of Mathematical Sciences, University College Cork, Western Road, Cork T12 XF62, Ireland
[*]chiara.poletto@inserm.fr

# 1 Methods

**Mean-field dynamics**

We studied the mean-field dynamical system described by equations (1) in the main paper by stability analysis and numerical integration. We derived closed-form expressions for all fixed-points and almost all conditions underlying their local stability. We used numerical evaluation of the Jacobian's spectrum to study stability whenever an analytical solution was not possible and to check the accuracy of the analytical results as well. We then numerically integrated the mean-field equations exploring different initial conditions. Numerical integration of the ordinary differential equations was performed in *Python 3.6* using the function *odeint* from the *Scipy* package.

## Mean-field dynamics with communities

The epidemic in each population follows the same dynamics as in equations (1) of the main text. The infection terms, however, must be modified in order to account for the different contributions (between and within community) to the force of infection. Specifically, the force of infection due to, e.g. $B_1$, acting on an individual in community $c = 1, 2$ becomes $\beta_1 \left[ (1-\epsilon) X_1^{(c)} + \epsilon X_1^{(c')} \right]$, where $c \neq c'$. The two distinct terms appearing in this expression represent the contributions due to infected individuals in the same community ($c$) and in the other community ($c'$), respectively. Notice that for 2 interacting populations, as considered in the main paper, we can reduce the number of independent equations from 12 to 10 by exploiting density conservation within each populations, i.e. $\sum_Z Z^{(c)} = 1$, where $Z^{(c)}$ denote the fraction of individuals in state $Z$ and community $c = 1, 2$. Numerical integration of these equations is performed in the same way as in the case of a single population.

## Network models

We used the algorithm outlined in [1] to efficiently generate Erdős-Rényi networks. In order to generate modular networks with $n_C$ communities and adjustable community strength, we first group nodes into $n_C$ different communities. Here we chose for simplicity to assign exactly $N/n_C$ nodes to each community. Each node receives a random number of open connections drawn from a Poisson distribution with average $\bar{k}$. We then classify each of these connections as either a within-community or an inter-community stub with probabilities $1 - \epsilon$ and $\epsilon$, respectively. Links are finally created by matching stubs. Within-community (between-community) stubs are matched with each other according to a configuration model. We eventually discard self-links, multiple links between any pair of nodes and unmatched stubs. For large networks, the number of discarded stubs is usually negligible compared to the number of links. Notice that this algorithm enables us to independently set both the degree distribution and the strength of the community structure.

## Simulating spread of concurrent diseases on networks

Simulations occur in discrete time. During each time step we check first for possible infection events caused by infected nodes and then for recovery events. Every infected node tries to transmit the disease(s) it is carrying to each of its neighbours. Each naive susceptible individual (compartment $S$) can get infected by pathogen $A$ with probability $1 - (1 - \alpha)^{n_A}$, where $n_A$ is the number of its susceptible neighbours carrying $A$. At the same time, a naive susceptible can also be infected by either $B_1$ or $B_2$. To avoid co-infections with $B_1$ and $B_2$, we loop over the neighbours of the naive node in a random order, checking for each infectious neighbour node if infection occurs or not (according to the corresponding infection probability, i.e. either $\beta_1$ or $\beta_2$) and stopping iteration at the first successful infection event. Transmission with either $B_1$ or $B_2$ can

occur independently from $A$, thus direct transitions from $S$ to either $D_1$ or $D_2$ compartments are allowed. Secondary infections are implemented in a similar way.

For convenience, the simulation time step $\Delta t$ was taken as the time unit. To avoid possible spurious effects due to time discretisation we set the infectious duration to be longer than the time step, i.e. $\mu^{-1} = 20\Delta t$. During each time step, infected individuals recover from each of the diseases they are carrying with probability $\mu$. As a consequence a doubly infected individual can turn into a fully susceptible individual with probability $\mu^2$. Individuals cannot recover during the same time step they got infected.

For each simulation, we initially choose 100 nodes for each pathogen and set them infected with that pathogen. In the Erdős-Rényi case initially infected nodes are chosen at random, whereas in the case of modular networks the infected seeds are chosen at random within the community where a particular pathogen is seeded. We stop simulations 400 time steps after either $B_1$ or $B_2$ becomes extinct, or, alternatively, after reaching the maximum simulation time $T_{max}$. The former stopping condition is dictated by the need to discern simulations where $A$ is able to persist from those where it becomes extinct right after extinction of either one of the $B$ strains. When any of the stopping conditions is met, we check which pathogens have survived and the corresponding prevalence. To reconstruct the phase diagram of figures 4 and 5 of the main paper we have 500 and 140 simulations for any given point in the parameter space for the Erdős-Rényi and modular networks respectively.

## 2 Results

### 2.1 Equilibria and stability analysis for the well-mixed system

Here we enumerate fixed points and study the stability of each equilibrium point by finding the eigenvalues of the corresponding Jacobian matrix $\mathcal{J}$. Because total density is conserved, the effective number of independent equations can be reduced from 6 to 5. Therefore $\mathcal{J}$ is a 5x5 matrix. In the following we eliminate $A$ exploiting total density conservation and consider $S, B_1, B_2, D_1$, and $D_2$ as independent variables. $A$ is kept as a placeholder for $1 - S - B_1 - B_2 - D_1 - D_2$. The general form of the Jacobian is given by:

$$\begin{pmatrix} \alpha(S^* - X_A^*) - 1 - \beta_1 X_1^* - \beta_2 X_2^* & S^*(\alpha - \beta_1) & S^*(\alpha - \beta_2) & -1 - \beta_1 S^* & -1 - \beta_2 S^* \\ c_1 \alpha B_1^* + \beta_1 X_1^* & c_1 \alpha(B_1^* - X_A^*) + \beta_1 S^* - 1 & c_1 \alpha B_1^* & 1 + \beta_1 S^* & 0 \\ c_2 \alpha B_2^* + \beta_2 X_2^* & c_2 \alpha B_2^* & c_2 \alpha(B_2^* - X_A^*) + \beta_2 S^* - 1 & 0 & 1 + \beta_2 S^* \\ -c_1(\alpha B_1^* + \beta_1 X_1^*) & c_1 \beta_1(A^* - X_1^*) + c_1 \alpha(X_A^* - B_1^*) & -c_1(\alpha B_1^* + \beta_1 X_1^*) & c_1 \beta_1(A^* - X_1^*) - 2 & -c_1 \beta_1 X_1^* \\ -c_2(\alpha B_2^* + \beta_2 X_2^*) & -c_2(\alpha B_2^* + \beta_2 X_2^*) & c_2 \beta_2(A^* - X_2^*) + c_2 \alpha(X_A^* - B_2^*) & -c_2 \beta_2 X_2^* & c_2 \beta_2(A^* - X_2^*) - 2 \end{pmatrix} \quad (1)$$

1. **Disease free state**: $S^* = 1$.

3

For this equilibrium $\mathcal{J}$ takes the form of an upper triangular matrix. Therefore the eigenvalues can be read off immediately since they coincide with the diagonal elements. In particular: $\lambda \in \{\alpha - 1, \beta_1 - 1, \beta_2 - 1, -2, -2\}$. Stability is therefore ensured if and only if all base transmission rates are smaller than 1.

2. **Pathogen A only** ($[A]$): $S^* = 1/\alpha, A^* = 1 - 1/\alpha$, which is feasible if and only if $\alpha > 1$.

   In this case one eigenvalue can be found immediately by inspection and it is equal to $1-\alpha$, which is always negative when this equilibrium is feasible (i.e. $\alpha > 1$). The rest of the Jacobian matrix takes a 2x2 block diagonal form with diagonal blocks $J_i^{[A]}$ ($i = 1, 2$) given by:

$$J_i^{[A]} = \begin{pmatrix} -1 + c_i(1-\alpha)\beta_i/\alpha & 1 + \beta_i/\alpha \\ c_i(\beta_i + \alpha)(1 - \alpha^{-1}) & -2 + c_i\beta_i/\alpha \end{pmatrix}, \quad (2)$$

   whose eigenvalues can be determined by considering the matrix $\Gamma(\lambda) = J_i^{[A]} - \lambda I_2$, where $I_n$ is the $n \times n$ identity matrix. We subtract the second column of $\Gamma$ from the first column obtaining a new matrix $\Gamma'$. By construction $\det(\Gamma) = \det(\Gamma')$. Therefore, the characteristic polynomial is the same. Now, however, the latter polynomial already appears in a factorized form, yielding the eigenvalues $\lambda = -2 - c_i(\alpha - 1)$, which is always negative, and $\lambda = -1 + \beta_i c_i(1 - \alpha^{-1}) + \beta/\alpha$, which is negative if and only if $\beta_i < \dfrac{\alpha}{1 + c_i(\alpha - 1)}$. For $\alpha > 1$ (condition for the solution to be feasible, as written above), this is never true if either one of $B_i$ is super-critical.

3. **Strain B$_i$ only** ($[B_i]$): $S^* = 1/\beta_i, B_i^* = 1 - 1/\beta_i$, which is feasible if and only if $\beta_i > 1$.

   In the following we will use the index $i$ to refer to strain $B_i$ while the index $j$ will indicate the competitor. Here $\mathcal{J}$ can be broken down into a 2x2 upper triangular matrix and a 3x3 matrix. The former has eigenvalues -2 and $-1 + \dfrac{\beta_j}{\beta_i}$. Therefore, stability requires $\beta_i > \beta_j$. The remaining 3x3 matrix $J_3^{(B_i)}$ takes the form:

$$J_3^{[B_i]} = \begin{pmatrix} \alpha S^* - 1 - \beta_i B_i^* & \alpha S^* - 1 & -2 \\ \alpha c_i B_i^* + \beta_i - 1 & \alpha c_i B_i^* & 2 \\ -c_i(\alpha + \beta_i)B_i^* & -c_i(\alpha + \beta_i)B_i^* & -2 - \beta_i c_i B_i^* \end{pmatrix}, \quad (3)$$

   which can be easily diagonalized by considering the matrix $\Gamma(\lambda) = J_3^{[B_i]} - \lambda I_3$ and performing the following row operations: first add its first row to its second row, then add the second row to its third row; the first row is left unchanged. This procedure enables writing down the characteristic polynomial in an easy-to-factorize form, yielding the eigenvalues $\lambda = 1 - \beta_i$, $\lambda = -2 - c_i(\beta_i - 1)$ and $\lambda = \alpha c_i B_i^* - 1 + \alpha \beta_i^{-1}$. The former two are always

4

negative when this equilibrium is feasible, while the latter is negative if and only if $\alpha < \dfrac{\beta_i}{1 + c_i(\beta_i - 1)}$, which is never true if the spreading of $A$ is super-critical.

4. **$A$ and $B_i$ syndemic ($[A\&B_i]$)**: By using the equilibrium conditions $\dot{X}_i = 0$ and $\dot{X}_A = 0$,
$$\beta_i(S^* + c_i A^*) - 1 = 0, \qquad (4)$$
and
$$\alpha(S^* + c_i B_i^*) - 1 = 0, \qquad (5)$$
we find that $B_i^* = \dfrac{1 - \alpha S^*}{c_i \alpha}$ and $A^* = \dfrac{1 - \beta_i S^*}{c_i \beta_i}$. By exploiting density conservation, i.e. $D_i^* = 1 - S^* - A^* - B_i^*$, we can express every variable in terms of $S^*$. The latter is determined by a quadratic polynomial whose roots are given by:

$$S_\pm^* = \dfrac{1 \pm \sqrt{1 - \dfrac{4}{c_i \beta_i \alpha}\left(1 - \dfrac{1}{c_i}\right)}}{2\left(1 - \dfrac{1}{c_i}\right)}, \qquad (6)$$

which exists if $\alpha \beta_i > 4(c_i - 1)/c_i^2$.

Here $\mathcal{J}$ can be broken down into a 2x2 and a 3x3 matrices. The smaller matrix $J_2^{[A\&B_i]}$ is given by:

$$J_2^{[A\&B_i]} = \begin{pmatrix} \beta_j S^* - \alpha c_j X_A^* - 1 & 1 + \beta_j S^* \\ -c_j(\alpha X_A^* + \beta_j A^*) & -2 - \beta_j c_j A^* \end{pmatrix}, \qquad (7)$$

which can be easily diagonalised by using row/column operations, yielding the eigenvalues $\lambda = -2 - c_j \alpha X_A^*$ and $\lambda = -1 + \beta_j(S^* + c_j A^*)$. The former is always negative while the latter corresponds to the asymptotic growth rate of $X_j$. This is the only eigenvalue in which parameters $\beta_j$, $c_j$, which pertain to the competitor strain, appear.

$J_3^{[A\&B_i]}$ instead takes the form:

$$\begin{pmatrix} \alpha(S^* - X_A^*) - \beta_i X_i^* - 1 & S^*(\alpha - \beta_i) & -1 - \beta_i S^* \\ c_i \alpha B_i^* + \beta_i X_i^* & c_i \alpha(B_i^* - X_A^*) - 1 + \beta_i S^* & 1 + \beta_i S^* \\ c_i(\alpha B_i^* + \beta_i X_i^*) & c_i \alpha(X_A^* - B_i^*) + c_i \beta_i(A^* - X_i^*) & -2 + c_i \beta_i(A^* - X_i^*) \end{pmatrix}. \qquad (8)$$

Although the spectrum of $J_3^{[A\&B_i]}$ cannot be determined analytically, we can still gain some insight about stability conditions by studying its characteristic polynomial. We do so by first computing $\Gamma(\lambda) = J_3^{[A\&B_i]} - \lambda I_3$. We then consider the matrix $\Gamma'$ obtained by adding the first row of $\Gamma$ to the second row and adding the second row to the third row. The characteristic equation takes the form $P(\lambda) = \lambda^3 + a_2 \lambda^2 + a_1 \lambda + a_0 = 0$. Now, the

5

Routh-Hurwitz criterion states that in order for all $P(\lambda)$'s roots to have a negative real part, the following conditions must be satisfied: $a_2 > 0$, $a_0 > 0$ and $a_2 a_1 > a_0$. We find that:

$$a_2 = 2 + (c_i + 1)(\alpha X_A^* + \beta_i X_i^*), \tag{9}$$

$$a_1 = \alpha \beta_i c_i^2 X_i^* X_A^* + c_i(2+\alpha X_A^*+\beta_i X_i^*)(\alpha X_A^*+\beta_i X_i^*)+(1-c_i)(\alpha^2 X_A^*+\beta_i^2 X_i^*)S^* \tag{10}$$

$$a_0 = \alpha \beta_i c_i^2 X_i^* X_A^* (\alpha + \beta_i)(1 - 2(1 - 1/c_i)S^*), \tag{11}$$

so $a_2 > 0$ always. Substituting $S_\pm^*$ inside the definition of $a_0$ yields $a_0(S_+^*) < 0$ and $a_0(S_-^*) > 0$. Therefore according to the Routh-Hurwitz criterion the solution $S_+^*$ is never found to be stable. $S_-^*$ is, instead, stable when the condition $a_2 a_1 > a_0$ is satisfied. The feasibility of $[A\&B_i]$ state can be checked numerically as well by requiring that $S^*, A, B_i^*, D_i^* > 0$. In particular, the condition $\alpha \beta_i > 4(c_i - 1)/c_i^2$ ensures that $S_-^*$ is real and positive and explains the vertical boundary delimiting the $[A\&B_2]$ stable region in figure 2b in the main manuscript.

5. **All strains coexist** We first consider the equilibrium conditions $\dot{X}_i = 0$, $i = 1, 2$, given by Eq. (4), which allow us to obtain $S^*$ and $A^*$:

$$S^* = \frac{c_2 \beta_2 - c_1 \beta_1}{\beta_1 \beta_2 (c_2 - c_1)},$$

$$A^* = \frac{\beta_2 - \beta_1}{\beta_1 \beta_2 (c_1 - c_2)}.$$

Notice that if $\beta_2 > \beta_1$, then we need $c_1 > c_2$ and $c_1 \beta_1 > c_2 \beta_2$ for this fixed point to be feasible. After some algebra one can obtain a quadratic equation for $B_1^*$:

$$\left(1 + \beta_2 S^*\right)\left(1 - S^* - A^*\right) + c_2^{-1}\left(S^* - \alpha^{-1}\right)\left(1 + \alpha S^* + \alpha c_2(1 - S^*)\right) +$$

$$B_1^* \left\{ -1 - \beta_2 S^* - \frac{1 + \beta_2 S^*}{1 + \beta_1 S^*}\left(1 - c_1/c_2 + \alpha c_1 - (\beta_1 + \alpha c_1(1 - c_2^{-1}))S^*\right) +\right.$$

$$\left(1 - \alpha S^*\right)\left(1 - c_1/c_2\right) + c_1/c_2\left(1 + \alpha S^* + \alpha c_2(1 - S^*)\right) \bigg\} +$$

$$B_1^{*2} \alpha c_1 \left(c_1/c_2 - 1\right)\left(1 - \frac{1 + \beta_2 S^*}{1 + \beta_1 S^*}\right) = 0. \tag{12}$$

6

Once the roots of the latter equations have been found, the remaining fractions can be computed using:

$$B_2^* = c_2^{-1}\Big(\alpha^{-1} - S^* - c_1 B_1^*\Big),$$

$$D_1^* = \frac{1 - \beta_1 S^* + \alpha c_1\Big(1 - S^* - B_1^* - B_2^*\Big)}{1 + \beta_1 S^*} B_1^*,$$

$$D_2^* = 1 - S^* - A^* - B_1^* - B_2^* - D_1^*.$$

By numerically computing these equilibria, we found that they can be feasible under certain conditions on the parameters. The condition for their stability cannot be computed analytically. However, we can still evaluate the Jacobian at the fixed points and then compute its eigenvalues numerically and find that the equilibria are always unstable.

**Additional results**

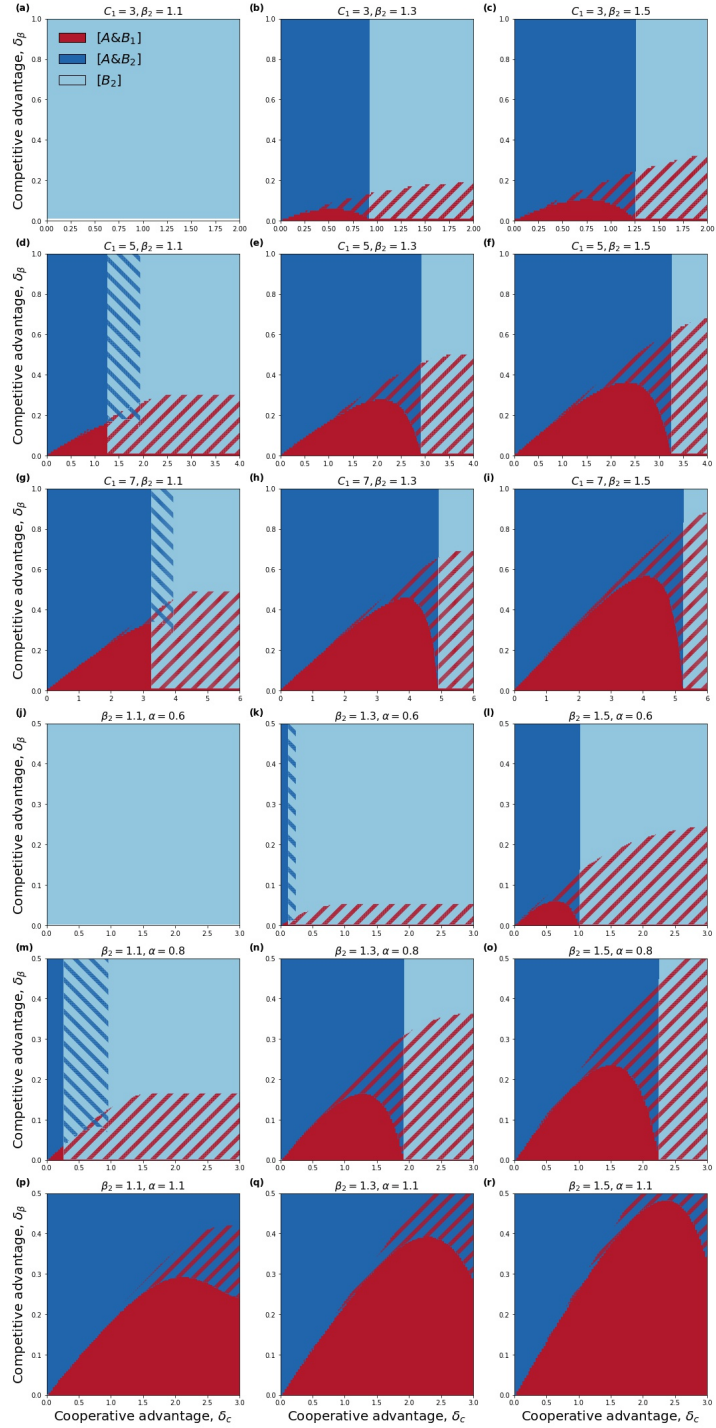

Figure S1: **Phase diagram for different values of $\beta_2$, $\alpha$ and $c_1$.** (a)-(i) $\alpha = 0.8$. (j)-(r) $c_1 = 4$. The range of $\beta_1$ includes values below 1.

8

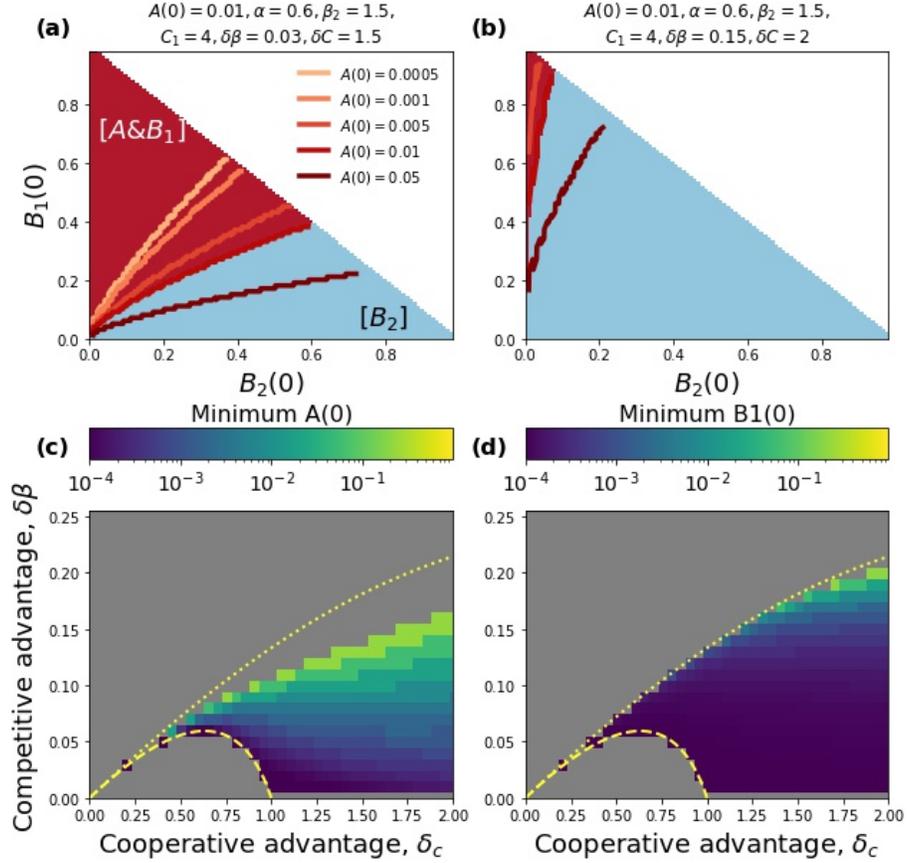

**Figure S2: Initial conditions and mean-field dynamics**. (a),(b) Final outcome as a function of $B_1(0)$ and $B_2(0)$ with $A(0) = 0.01$. Additional boundaries separating the two dominance regions and corresponding to different values of $A(0)$ are also showed. For each value of $A(0)$ we explore values of $B_1(0)$ and $B_2(0)$ in the simplex $0 \leq B_1(0) + B_2(0) \leq 1 - A(0)$. Prameters were: (a) $\delta_\beta = 0.03$, $\delta_c = 1.5$, (b) $\delta_\beta = 0.15$, $\delta_c = 2$. (c) Minimum amount of $A(0)$ required in order for $B_1$ to win as a function of $\delta_\beta$ and $\delta_c$, given that $B_1(0) = B_2(0) = 0.001$. (d) Minimum amount of $B_1(0)$ required in order for $B_1$ to win as a function of $\delta_\beta$ and $\delta_c$, given that $A(0) = 0.01$ and $B_2(0) = 0.001$. Grey regions in (c,d) correspond to either absence of bistability or to parameter choices for which $B_1$ never wins. This figure shows that a sufficiently large advantage in terms of initial conditions can lead to $B_1$ winning the competition in the bistable region. Other parameters are the same as in figure 2a of the main manuscript.

9

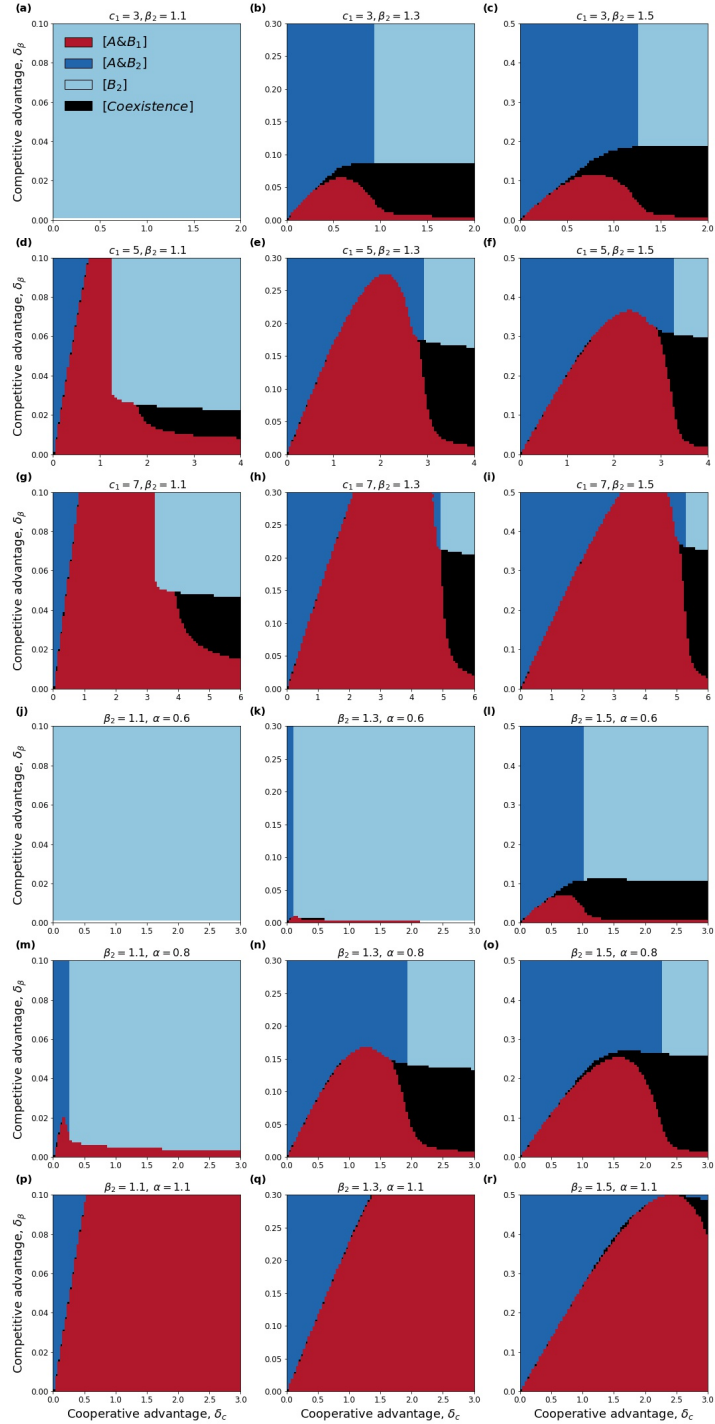

Figure S3: **Equilibrium configuration for two interacting communities for different values of $\beta_2$, $\alpha$ and $c_1$.** (a)-(i) $\alpha = 0.8$. (j)-(r) $c_1 = 4$. Initial conditions are as in figure 4c of the main paper. Here we have considered values of $\delta_\beta$ and $\delta_c$ such that $\beta_1$, $\beta_2$, $c_1$ and $c_2$ are all larger than unity. Increasing values of $\alpha$ have a positive effect on the persistence of $B_1$.

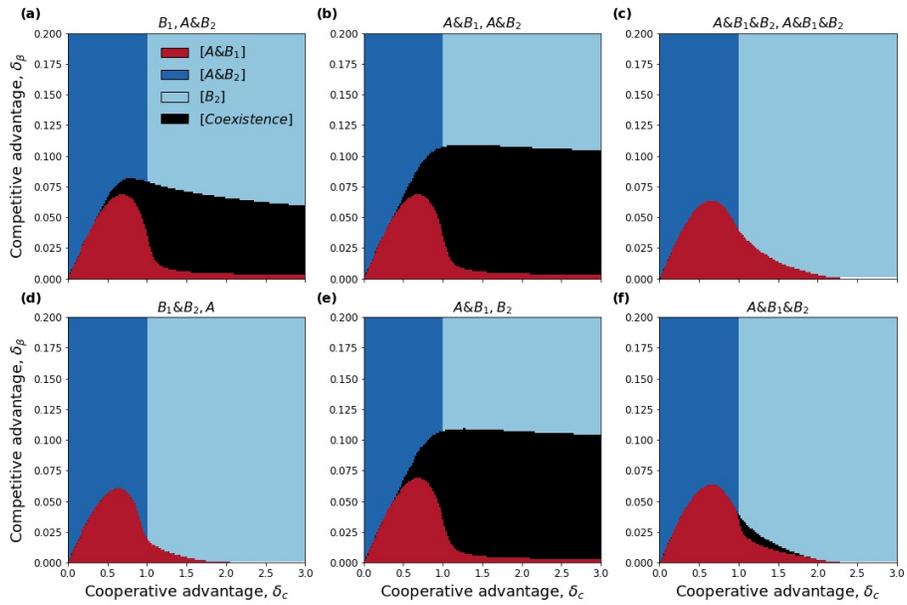

**Figure S4: Role of initial conditions in mean-field dynamics with communities.** Final outcome as a function of $\delta_c$ and $\delta_\beta$ for different initial conditions. The title of each panel indicates in which communities each pathogen is seeded into; the initial density of a pathogen in the community it is seeded into is set to 0.01. Other parameters are as in figure 4 of the main manuscript.

11

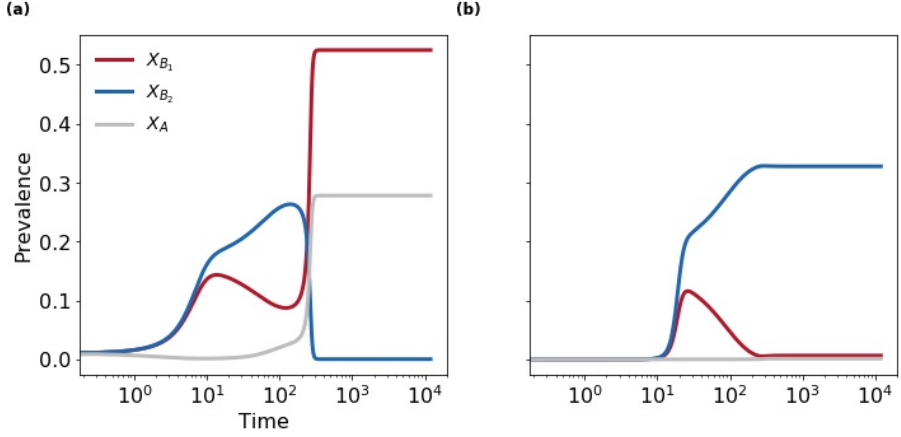

**Figure S5: Dynamics with two strains seeded in the same community.** Dynamical trajectories of $B_1$'s, $B_2$'s and $A$'s total prevalence (in red, blue and gray respectively) in community 1 (a) and 2 (b). Trajectories were obtained numerically by starting from $B_1^{(1)}(t=0) = B_2^{(1)}(t=0) = A^{(1)}(t=0) = 0.01$. Around $t = 10$, $B_2$ takes over community 2 as a results of its advantage in transmissibility. At this point $B_1$'s prevalence declines until $A$ gives rise to a new outbreak; the latter ultimately leads to a $B_1$ dominance in the first community. Here we set $\delta_\beta = 0.025$, $\delta_c = 1.2$ and $\epsilon = 0.0002$. Other parameters are as in figure 2a.

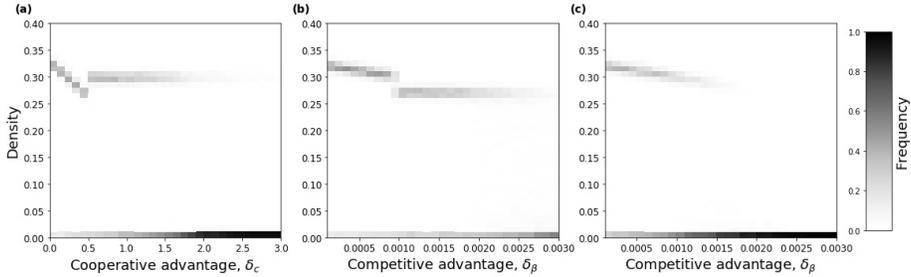

**Figure S6: Final-state density of $X_A$ for stochastic simulations on ER networks.** (a-c) show the probability of the final state for a given value of $X_A$. Here we do not differentiate between which final state is attained, i.e. which pathogen persists. (a) has been obtained for $\delta_\beta = 0.001$, while (b) and (c) have been obtained for $\delta_c = 0.4$ and $\delta_\beta = 1.5$, respectively. In (a) we can observe two types of transitions in the behavior of $X_A$: the sharp transition at $\delta_c = 0.5$ corresponds to a leap from state $[A\&B_2]$ to state $[A\&B_1]$. As $\delta_c$ is further increased, extinction of $A$ becomes increasingly probable as the state $[B_2]$ is reached with a higher frequency. This transition is, however, gradual and around $\delta_c = 1.3$ a crossover is reached where states $[A\&B_1]$ and $[B_2]$ are reached with equal probability. Other parameters are as in figure 6.



\end{document}